\documentclass[letterpaper,UKenglish]{llncs}
\usepackage{amsfonts}
\usepackage{amsmath}
\usepackage{amssymb}
\usepackage{hyperref}
\usepackage{algorithm}
\usepackage{algpseudocode}
\usepackage{algorithmicx}
\usepackage{float}
\usepackage{microtype}

\usepackage{tikz}
\usetikzlibrary{automata,positioning, arrows}

\usepackage{bbm}

\title{Random Words in a (Weighted) Regular Language: a Free Energy Approach}

\author{Cewei Cui and Zhe Dang}
\institute{School of Electrical  Engineering and Computer Science\\
Washington State University, Pullman, WA 99164,  USA\\
  \texttt{\{ccui,zdang\}@eecs.wsu.edu}}
\authorrunning{C.\, Cui and Z.\, Dang} 

\begin{document}

\maketitle

\begin{abstract}
We study random words in a weighted regular language  that achieve the maximal free energy using thermodynamics formalism. In particular, typical words in the language are algorithmically generated which have applications in
computer security (anomaly detection) and software enegineering (test case generation).
 \end{abstract}

\section{Introduction}
A random walk is essentially a stochastic process, which has been widely used
in many disciplines, including mathematics, physics, and finance.
A Markov chain is probably the simplest and yet most useful form of random walks,
where its ergodicity, hitting time, convergence time etc. have all been well-studied
\cite{lovasz1993random,aldous2002reversible}.
On the other hand, many physical behaviors can be modeled by random walk models, such as Brownian motion,
and Ising model. In finance, a famous book ``A random walk down wall street" \cite{malkiel1999random}
shed light to the path of using mathematical
modelling to predict the stock market.

Random walks have applications in Computer Science as well, in particular
in using random walks on graph data structures (such as graph represented image and social network). Turning to automata theory,
a wide spectrum of applications can also be found, from
the earliest  Rabin automata   (dated back to 1960s)
 all the way to more modern quantum automata.
Since automata are the fundamental model for all computer programs (e.g., an instruction can be understood as a transition), this line of research addresses the problem
 on, given transition probabilities in a computer program,  how the program is going to behave. This is important, since such programs are in turn  the backbone of
randomized algorithms.
However, what we are looking for here concerns
 a related but essentially different problem:

Given: a program (without probabilities),

Question: what would be the program's ``most random" behaviors?

\noindent Clearly, one can re-interpret the problem in various contexts. For instance,

Given: a regular language (with no probabilities),

Question: what are the ``most random" words?

\noindent As we shall state later, answers to these questions have applications in
many areas in Computer Science (such as in network intrusion detection,
one would like to know
what the normal/abnormal traffic is, without seeking help from some pre-given probabilistic model -- in reality, properly assigning probabilities in a behavioral model is
difficult and of course error-prone).

These problems are not new at all, where their most common
solutions require computation of a random walk (since it is not given)
on a graph using,
 to our best knowledge,  one of the following
 two main approaches:
\begin{enumerate}
\item a symmetric random walk on a graph
with uniform probability assignment on branches from the same node;
\item a maximal entropy random walk on a graph to achieve roughly
uniform probability on paths.
\end{enumerate}
 Turning back to automata theory,
the latter approach has been followed by, e.g.,
Basset's maximal entropy stochastic process on timed automata \cite{basset2015maximal}.
The random walk model computed
in our research is different from the aforementioned two approaches, since we
focus on
weighted graphs, where a weight can model cost, time, etc for an event.
 Simply speaking,
the computed
probability of a path  is proportional to its weight. If all paths
with the same length
 take the same weight,
it degenerates to the maximal entropy model. In practice, a
 weight can be used to model time, cost, or  risk level, etc., associated with an event.

The theoretical underpinning of our approach comes from
thermodynamic formalism \cite{ruelle2004}, where
a weighted regular language (in this paper, we consider the simplest form weighted language where each symbol in the alphabet is given a weight) is the collection
of trajectories of a gas molecule moving along the weighted graph defining the minimal
deterministic finite automaton (DFA) accepting the language.
When the graph is strongly connected, one can compute a random walk -- a unique Markov
chain $\mu^*$ -- that achieves the maximal free energy of the gas molecule, using
the theory developed by Gurevich in  the 80's.
Clearly, from the computed $\mu^*$, one can generate random words
(we call them Gurevich random words)
in the weighted language.
Hence,
  we call our approach
 a free energy one.

 The rest of the paper is organized as follows.
In Section 2, we present the basics physics
behind Gurevich free energy.
In Section 3, Gurevich random words in weighted
regular languages are proposed and investigated.
In Section 4, we study the AEP (Asymptotic Equipartition Property)
in  a (weighted) regular language and hence typical words in the language can then be
computed.
In Section 5, we discuss the applications of the random words
 in software testing and computer security.
We conclude  the paper in Section 6.

\section{Physics of Gurevich Free Energy}
When a gas molecule moves, measurements such as its position, speed, etc. can be made at various times.
Each such measurement is called a micro-state and thus, we obtain a sequence, called an orbit,
 of micro-states
$\alpha_n=s_0,\cdots, s_n$. There is a change
in potential  $w(s_i,s_{i+1})$ when the molecule moves from $s_i$ to
$s_{i+1}$ and therefore,
the total potential change on the moves from $s_0$ all the way to $s_n$ is clearly
\begin{equation}
w(\alpha_n)=\sum_{0\le i\le n-1} w(s_i,s_{i+1}).
\end{equation}
The Boltzmann equation defines the relationship between energy  and choice:
$$E=k_{\rm B}\ln {\cal C},$$
where the energy $E$ of the molecule can be expressed as the logarithm of the number of choices
(of the next micro-state), where $k_{\rm B}$ is the Boltzmann constant.
This is not hard to understand at all: an active molecule carries high energy since it has more choices to move around.
Therefore, $e^{w(\alpha_n)}$ (we ignore the Boltzmann constant)
 will be the total number of choices in the process when the molecule moves on the orbit $\alpha_n$ that is
from
$s_0$ to $s_n$.  Summing up the choices for all the orbits $\alpha_n$ and then taking logarithm, we will obtain the energy of the molecule
on orbits with length $n$:
$$\ln\sum_{\rm all ~\alpha_n}  e^{w(\alpha_n)}.$$
However, this would not give a finite value as $n\to\infty$. We now consider the asymptotic energy (per step)
\begin{equation}\label{GU}
\lambda=\limsup_{n\to\infty}{1\over n}\ln\sum_{\rm all ~\alpha_n}  e^{w(\alpha_n)}.
\end{equation}
Unfortunately, to compute the $\lambda$ is not trivial at all and has been one of the central topics in thermodynamic formalism \cite{ruelle2004}. A crowning achievement in the decades of research in the formalism is the following Variational Principle \cite{ruelle2004} (with some side conditions, which are ignored for now for simplicity of our presentation):
\begin{equation} \label{varprinciple}
\limsup_{n\to\infty}{1\over n}\ln\sum_{\rm all ~\alpha_n}  e^{w(\alpha_n)}=\sup_\mu (\int w d\mu + H_\mu),
\end{equation}
which can be intuitively understood as the following:
\begin{itemize}
\item The LHS, which is the aforementioned energy $\lambda$, can be computed through the RHS;
\item To compute the RHS, one would find a $\mu$, a discrete time Markov chain,
 that maximizes the sum of the average potential change per step and the Kolmogorov-Sinai entropy.
\end{itemize}
A seminal result is due to Gurevich \cite{Gurevich1984}: the RHS can indeed be
achievable by a unique Markov chain
$\mu^*$ when the $w(\cdot,\cdot)$ is defined on a finite graph
(i.e., the aforementioned micro-state is a node in the graph) that is an SCC (strongly connected component).
The Markov chain
$\mu^*$
 has a unique stationary distribution and assigns every edge of the graph with a transition probability. The RHS is called Gurevich free energy and thus, we call the $\lambda$ as the free energy of
 the molecule. We now sketch the Gurevich algorithm in computing the $\lambda$ and the $\mu^*$.

Let $G$ be a (directed) graph with nodes in $V$ and edges in $E$.
We consider a gas molecule's moving as a walk on $G$ while a node
in $V$ resembles a micro-state of the molecule.
And hence, the gas molecule can be observed as a sequence of nodes; i.e.,
a walk in $G$ (herein, a walk may contain loops). In particular when an edge $e$ in $E$ is associated with a weight
$w({e}) \in {\bf Q}$(rationals), the weight resembles the potential.
Notice that, as a special case, if the $G$ is not weighted graph, we may
simply take $w({e})=0$ for all ${ e} \in E$.
We shall note that there are no probability assignments on the edge
of $E$ and hence $G$ is just a graph.

Given these, how can we create the ``most random" walks on $G$?
In our opinion, the walk that we are looking for shall maximize the free
energy among all possible walks on $G$; i.e., the Gurevich
$\mu^{*}$ that
achieves the superume on the RHS of the aforementioned Variational Principle.
The way to compute $\mu^{*}$  is laid out in the seminal paper
of Gurevich\cite{Gurevich1984} where the graph $G$ is required to be strongly
connected, and is shown in Algorithm \ref{Gurevich-algorithm}.
 Notice that the free energy rate $\lambda$ in (\ref{GU}) can also be computed  as $\lambda=\ln\hat\lambda$ where
 $\hat\lambda$ is the Perron number computed  in the algorithm.

\begin{algorithm}[H]
\begin{algorithmic}[1]

\footnotesize

\Require{${\cal M}$  is the adjacency matrix of a strongly connected  weighted graph $G=<V,E>$.
Each entry in ${\cal M}$, ${\cal M}_{i,j}$, represents the weight from
node $i$ to node $j$ in $G$.
If there is no edge from node $i$ to node $j$ in $G$,
we simply take the ${\cal M}_{i,j}=-\infty$.}

\Statex
\Function{ComputeTransitionProbability}{${\cal M}$}
 \State First build a weight matrix $W=\{W_{i,j}\}$
 \For {$i$, $j$ in \{ ${\cal M}_{ij}$\}}
    \If{${\cal M}_{i,j}\ne -\infty$}
        \State $W_{i,j} = e^{{\cal M}_{i,j}}$
    \Else
        \State $W_{i,j} = 0$
    \EndIf
 \EndFor
 \State Conduct eigen decomposition on the matrix $W$.
 \State Obtain the right eigenvector of matrix $W$, $v=(v_1,v_2, \dots, v_n)$.
 \State Compute the  Perron number (i.e., the largest eigenvalue) of $W$, $\hat\lambda$.
 \State Using Parry measure,  obtain
 the transition probability matrix $P=\{P_{i,j}\}$ that defines the $\mu^*$:
 \For {$i$, $j$} in $W_{i,j}$
 \State $P_{ij} = \frac{W_{i,j} v_j}{\lambda v_i}$
 \EndFor
  \EndFunction

  \end{algorithmic}

 \caption{Computing Gurevich random walk from a weighted graph that is strongly connected}
 \label{Gurevich-algorithm}
 \end{algorithm}

\section{Gurevich Random Words in a (Weighted) Regular Language}\label{section2}

Let $L$ be a regular language on alphabet $\Sigma$. In particular,
each symbol $a\in\Sigma$
 is associated with a weight $w(a)\in \bf Q$ (rationals).
 Our random word generation uses Algorithm \ref{Gurevich-algorithm} (hence we call them Gurevich random words)
and therefore relies on a weighted graph. Herein, the graph is a DFA
accepting $L$.

Let $M$ be a DFA on alphabet $\Sigma$ and with states in
 $Q$ where $q_0$ is the initial state
and $F \subseteq Q$ is the set of accepting state. In particular, each symbol $a\in\Sigma$ carries the aforementioned weight $w(a)$.
We use $p \xrightarrow{a} q$ to denote  the
transition from state $p$ to state $q$ while reading input symbol $a$ with weight
$w(a)$. Many times we  write $p \xrightarrow{a,w(a)} q$ when we want to emphasize the weight $w(a)$.

Now, consider a run of $M$ from the initial state, for some $n>0$,
$$ q_0 \xrightarrow{a_1} q_1 \xrightarrow{a_2} q_2 \xrightarrow{} \dots \xrightarrow{a_n} q_n,$$
where each $q_{i-1} \xrightarrow{a_i} q_i$ is a transition in $M$.
The run is called the run on the word $\alpha= a_1 a_2 \dots a_n$.
As usual, when $q_n$ is an accepting state in the run, we say that $M$ accepts the word $\alpha=a_1 a_2 \dots a_n$.
 We use $L(M)$ to denote all words $\alpha$ accepted by $M$.

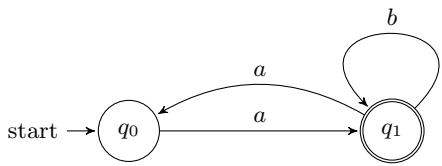
\begin{figure}
\begin{tikzpicture}[->,>=stealth',shorten >=1pt,auto,node distance=3.5cm,
        scale = 1,transform shape]

  \node[state,initial] (q0) {$q_0$};
  \node[state, accepting] (q1) [right of=q0] {$q_1$};

  \path (q0) edge              node {$a$} (q1)
        (q1) edge    [bend right, above]          node {$a$} (q0)
        (q1) edge    [loop, above]          node {$b$} (q1);
\end{tikzpicture}
\caption{An example DFA with weights $w(a)=2, w(b)=-1$.}\label{exdfa}
\end{figure}

We assume that $M$ is cleaned up. That is, every state in $M$ is reachable
from the initial state and every state can reach an accepting state
(and hence we do not consider the trivial case when $M$ accepts the empty language).
An example DFA $M$ is shown in Figure \ref{exdfa}.

\begin{figure}
\begin{tikzpicture}[->,>=stealth',shorten >=1pt,auto,node distance=3.5cm,
        scale = 1,transform shape]

  \node[state,initial] (q0) {$q_0$};
  \node[state, accepting] (q1) [right of=q0] {$q_1$};

  \path (q0) edge              node {pr=1, $a$} (q1)
        (q1) edge    [bend right, above]          node {pr=0.5, $a$} (q0)
        (q1) edge    [loop, above]          node {pr=0.5, $b$} (q1);
\end{tikzpicture}
\caption{Uniform probability assignments for the example DFA.}\label{exdfa1}
\end{figure}
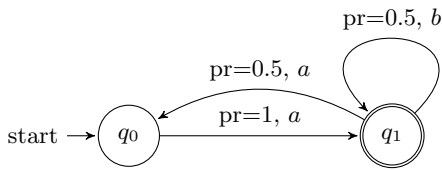

 How to generate a random word in
$L(M)?$ Of course, the randomness shall depend on the weights $w(\alpha)$ assigned to symbols $a$ in
$\Sigma$.
A straightforward way to obtain such random words would be use a most ``natural" way to assign
transition probabilities on transitions in $M$, each such probability is uniform among all outgoing
transitions from each state. For instance,  the probability assignments shown in Figure \ref{exdfa1} for the DFA in
Figure \ref{exdfa}.
However, there are problems with such uniform probability assignments:
\begin{itemize}
\item Though there could be many DFA $M$'s that accept the same $L$, the resulting
 probability assignments are not the same.
Hence, the ``randomness" depends on the choice of $M$, instead of $L$.
(2). The probability  assignments must be associated with the weights in certain ways such that
the resulting ``randomness" conforms with established randomness metrics like Shannon's
information rate $\lambda_L$ \cite{shabook}  (see also our recent work \cite{CuiDFI16,CuiDFI17})
 of the $L$, under the special case when all weights are 0 (unweighted words).
Herein, the rate $\lambda_L$ is defined (by Shannon) as
$$\lambda_L=\limsup_{n\to\infty} {{\log |L|_n}\over n},$$
where ${|L|_n}$ is the number of words with length $n$ in $L$. For instance,
if $w(a)=w(b)=0$ in the example DFA $M$ shown above, we can compute the information rate
of the regular language $L=a(aa+b)^*$ accepted by the $M$ is $\lambda_L=0.4812$.
However, the entropy rate of the Markov chain as the result of uniform probability assignments, as shown in
Figure \ref{exdfa1}, is $0.4621$. The two rates do not conform.
\end{itemize}

To address the problems, we first present an algorithm to generate random words in $L$
based on Gurevich's Algorithm \ref{Gurevich-algorithm}, where the generated random words achieve the
maximal free energy. We considered it as a most natural way to generate random words since, if
a word in $L$ reflects a potential change between measurements in a gas molecule, then the nature tends to
make the molecule to be in maximal free energy (actually, in the physical world, the free energy shall be minimal -- this is due to the fact that the sign of the energy is flipped in thermodynamics formalism for mathematical convenience, see more details in Sarig's notes \cite{sarig-notes}).

Let $M$ be the minimal DFA that accepts $L$. Notice that $M$ is cleaned up as we have mentioned earlier.
It is known that the $M$ is
unique (up to isomorphism on state names)
\cite{HopcroftU79}.
Applying the weight function $w$, the DFA $M$ is also weighted where
each transition with input symbol $a$ is assigned weight $w(a)$.

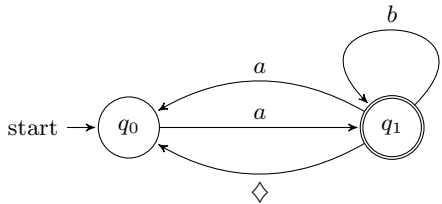
\begin{figure}
\begin{tikzpicture}[->,>=stealth',shorten >=1pt,auto,node distance=3.5cm,
        scale = 1,transform shape]

  \node[state,initial] (q0) {$q_0$};
  \node[state, accepting] (q1) [right of=q0] {$q_1$};


  \path (q0) edge              node {$a$} (q1)
        (q1) edge    [bend right, above]          node {$a$} (q0)
        (q1) edge    [loop, above]          node {$b$} (q1)
       (q1) edge     [bend left]    node {$\diamondsuit$} (q0);
\end{tikzpicture}
\caption{The $M_\diamondsuit$ for the DFA in Figure \ref{exdfa}.}\label{exdfa2}
\end{figure}

We first make the $M$ be strongly connected.
Let $\diamondsuit\not\in\Sigma$ be a new symbol.
For each accepting state $q$, we add
a transition to the initial state $q_0$:
$$q \xrightarrow{\diamondsuit} q_0.$$
The resulting DFA is written $M_\diamondsuit$.
The example automaton in Figure \ref{exdfa} is now modified into the automaton in Figure \ref{exdfa2}.

One can verify that $M_\diamondsuit$ is indeed strongly connected.
(Note that for the example automaton in Figure \ref{exdfa}, it is already strongly connected.
For such an automaton $M$, there is no need in converting them into $M_\diamondsuit$. For the ease of presentation,
we convert it anyway.)  The new symbol carries a weight $w(\diamondsuit)$ that is a pre-given very small
negative number (i.e., $\ll$ the minimal weight on $M$). We call these $\diamondsuit$-transitions as $\hat\epsilon$-edges while we bear in mind that $\hat\epsilon=e^{w(\diamondsuit)}$ is a positive number very close to 0.
(Adding such $\hat\epsilon$-edges to make a graph strongly connected  is not a new idea; it has
been used in the Google page rank algorithm \cite{page1999pagerank}.)

\begin{figure}
\begin{tikzpicture}[->,>=stealth',shorten >=1pt,auto,node distance=2.1cm,
        scale = 1,transform shape]

  \node[state] (q0) {$q_0$};
 \node[state] (q01) [right of=q0] { $q_{0a1}$};
  \node[state] (q1) [right of=q01] {$q_1$};
\node[state] (q10) [above of=q01] {$q_{1a0}$};
\node[state] (q11) [above of=q1] {$q_{1b1}$};
  \node[state] (q1-diamondsuit) [below of=q01] {$q_{1\diamondsuit0}$};

  \path (q0) edge              node {$w(a)$} (q01)
         (q01) edge              node {$0$} (q1)
        (q1) edge           node {$w(a)$} (q10)
(q10) edge              node {0} (q0)
        (q1) edge    [bend left]          node {$w(b)$} (q11)
  (q11) edge    [bend left]          node {$0$} (q1)

       (q1) edge         node {$-1000$} (q1-diamondsuit)
(q1-diamondsuit) edge           node {$0$} (q0);
\end{tikzpicture}
\caption{The weighted graph $G_M$ converted from $M_\diamondsuit$ in Figure \ref{exdfa2}, with $w(\diamondsuit)=-1000$.}\label{exdfa3}
\end{figure}
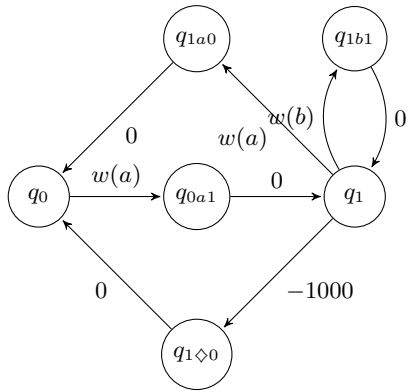

Next, we convert $M_\diamondsuit$ into a graph $G_M$ as follows. Notice that DFA $M$ itself may not be a graph:
there could be multiple transitions from a node to another.
Each state in $M_\diamondsuit$ is also a node in $G_M$. Additionally, $G_M$ has some other nodes shown below.
Initially, $G_M$ has no edges.
For each transition $p\xrightarrow{a} q$, with $a\in\Sigma\cup\{\diamondsuit\}$,
 in $M_\diamondsuit$, we add a new node
$\langle p, a, q\rangle$ and add the following two edges to $G_M$:
\begin{itemize}
\item the edge, with weight $w(a)$,  from node $p$ to node $\langle p, a, q\rangle$;
\item the edge, with weight 0,  from node $\langle p, a, q\rangle$ to node q.
\end{itemize}
It is not hard to verify that the resulting weighted graph $G_M$ is strongly connected.
Figure \ref{exdfa3} shows the result of $G_M$ from the $M_\diamondsuit$ in Figure \ref{exdfa2}.

\begin{figure}
\begin{tikzpicture}[->,>=stealth',shorten >=1pt,auto,node distance=3.5cm,
        scale = 1,transform shape]

  \node[state,initial] (q0) {$q_0$};
  \node[state, accepting] (q1) [right of=q0] {$q_1$};

  \path (q0) edge              node {pr=1.0000, $a$} (q1)
        (q1) edge    [bend right, above]          node {pr=0.9514, $a$} (q0)
 (q1) edge    [bend left, below]          node {pr=0.0000, $\diamondsuit$} (q0)
        (q1) edge    [loop, above]          node {pr=0.0486, $b$} (q1);
\end{tikzpicture}
\caption{The resulting probabilistic DFA $\hat M$ for the example DFA $M$ in Figure \ref{exdfa}.}\label{exdfa4}
\end{figure}
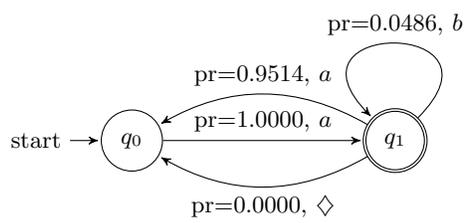

\noindent We then run Algorithm \ref{Gurevich-algorithm} on the graph $G_M$.
As a result, we obtain transition probabilities on each edge of $G_M$.
In particular, suppose that
$\theta$ is the transition probability computed
 on the edge from node $p$ to node $\langle p, a, q\rangle$, with $a\in\Sigma\cup\{\diamondsuit\}$,
we now label a transition probability $\theta$ to the transition $p\xrightarrow{a}q$
in the original DFA $M_\diamondsuit$.  As a result, we obtain a probabilistic DFA  $\hat M$ where each transition
is assigned with a probability.  Figure \ref{exdfa4} shows the resulting $\hat M$
after we run the Algorithm \ref{Gurevich-algorithm} on the graph $G_M$ shown in Figure \ref{exdfa3}.
One may compare the different probability assignments in Figure \ref{exdfa4} and in Figure \ref{GU}.

Notice that, because of the introduction of additional $\diamondsuit$-transitions,  the resulting $\hat M$ may not be
a perfect Markov chain (after dropping the  $\diamondsuit$-transitions).
In other words, for an accepting state, the sum of all transition probabilities on all
the outgoing transitions from the accepting state could be strictly less than 1 but still
very close to 1 because of the choice $e^w(\diamondsuit)$ to be very close to 0 (see Figure \ref{exdfa4} for an example).  We of course can tolerate this, since $\hat M$ itself already presents a random algorithm to
generate random words from the $L$:

\begin{algorithm}[H]
\begin{algorithmic}[1]

\footnotesize

\Require{$P=\{P_{i,j}\}$ is the transition probability matrix of ${\hat M}$,
and $E=\{E_{i,j}\}$} is the transition set of ${\hat M}$

\Function{GenerateRandomWord}{$P$}
 \State For every state in ${\hat M}$, build its cumulative transition probability matrix.
 \For {$q_i$ in ${\hat M}$}
    \State Use $P^{cum}_{i,j}$ to denote its cumulative transition probability.
    \State $P^{cum}_{i,j}=P_{i,0} + P_{i,1} + \dots + P_{i,j}$
    \State $P^{cum}_{i,-1} =0$
 \EndFor
 \State Let $q_x = q_{init}$ and $n=0$
 \State Generate a random number $r \in (0,1]$
 \For {$j$ in $\{P^{cum}_{x,j}\}$}
    \If  {$ r > P^{cum}_{x,j-1}$ and $r \le P^{cum}_{x,j}$}
        \State Output the symbol on edge $E_{i,j}$
        \State $n++$
        \State Let $q_x = q_j$ break from the loop.
    \EndIf
 \EndFor
 \State Goto 9 or if $q_x$ is an accepting state and $n>N$ Stop.
  \EndFunction

  \end{algorithmic}

 \caption{Generate a random word on ${\hat M}$ with a given lower bound $N$ on walk}
 \label{GenerateWord-algorithm}
 \end{algorithm}

We use Fig. \ref{exdfa4} as an example to explain the Algorithm \ref{GenerateWord-algorithm}.
Let $q_0$ be the initial state. Using the algorithm we have $P^{cum}_{0,0}=0$  $P^{cum}_{0,1}=1.0$,
$P^{cum}_{1,0}=0.9514$, $P^{cum}_{1,1}=1.0$.
The following is an example run of the algorithm.
\begin{itemize}
    \item Current state is $q_0$; generate a random number $r=0.8$
    \item Output the symbol $a$, because $0.8> P^{cum}_{0,0}$ and $0.8 \le P^{cum}_{0,1}$;
        Change the current state = $q_1$
    \item Current state is $q_1$; generate a random number $r=0.3$.
    \item Output the symbol $a$, because $0.3> P^{cum}_{1,-1}$ and $0.3 \le P^{cum}_{1,0}$.
            Change the current state = $q_0$.
    \item \dots
    \item Current state is $q_1$; generate a random number $r=0.97$.
    \item Output the symbol $b$, because $0.97> P^{cum}_{1,0}$ and $0.97 \le P^{cum}_{1,1}$.
            Change the current state = $q_1$.
    \item \dots
\end{itemize}

We shall point out that the random words generated by Algorithm \ref{GenerateWord-algorithm}
is a word   in $L^\diamondsuit=(L\diamondsuit)^*L$; i.e., a sequence of words in
$L$ separated by the $\diamondsuit$'s.  For practical purposes, each such word in $L$
 can be treated as a
Gurevich random words in $L$. Notice also that this treatment of the random words in $L$ will
make it possible to generate Gurevich random words in a finite language $L$ (whose  rate, by definition,
is $-\infty$).

 In the following, we will prove that our algorithm of the random walk is ``correct" in the sense that the rate $\lambda_\diamondsuit$
of the weighted regular language accepted by
$M_\diamondsuit$ approaches to the rate $\lambda$ of the weighted regular language $L$ as the weight $w(\diamondsuit)$
assigned on the $\hat\epsilon$-edges goes to $-\infty$:
\begin{equation}\label{good}
\lambda_\diamondsuit\to\lambda {\rm ~as~} w(\diamondsuit)\to-\infty.
\end{equation}
We first need clearly define
what the rates are in below:
\begin{equation}\label{deflambda}
\lambda =\limsup_{n\to\infty} {1\over n}\ln\sum_{\alpha\in L, |\alpha|=n} e^{w(\alpha)},
\end{equation}
and
\begin{equation}\label{defdimond}
\lambda_\diamondsuit =\limsup_{n\to\infty} {1\over n}\ln\sum_{\alpha\in L^\diamondsuit, |\alpha|=n} e^{w(\alpha)},
\end{equation}
where $L^\diamondsuit=(L\diamondsuit)^*L$ is the language accepted by $M_\diamondsuit$.
Recall that we run the Algorithm \ref{Gurevich-algorithm} on the graph $G_M$ constructed from $M_\diamondsuit$, so that we finally assign probabilities on the
transitions in $\hat M$. Notice that, inside the algorithm, the Perron number $\hat\lambda$ is also computed. The main result of our
earlier paper \cite{CuiD16} (the energy rate of a regular language can be computed  as
the energy rate of the DFA (as a graph), when the graph is strongly connected) shows that
\begin{equation}\label{earlier}
\lambda_\diamondsuit=2\ln\hat\lambda.
\end{equation}
We shall now thread this earlier result together with the Variational Principle:
the probabilistic program we obtained on  $G_M$, which is a strongly connected
graph without initial and accepting nodes, defines the $\mu^*$ that achieves
the RHS of the principle, due to Gurevich \cite{Gurevich1984}:
 \begin{equation}
\ln\hat \lambda=\limsup_{n\to\infty} {1\over n}\ln\sum_{\substack{
\alpha{\rm ~is~a~walk~on~}G_M\\ {\rm~with~length~}n} }
e^{w(\alpha)}=\int wd\mu^*+H_{\mu^*},
\end{equation}
where the walk $\alpha$ can be between any two nodes.
That is, the random walk $\mu^*$ on $G_M$, as the result of the  Algorithm \ref{Gurevich-algorithm}, does achieve the rate $\ln\hat\lambda$.
What's the difference between the random walk in $\hat M$ and the random walk
in $G_M$? Each transition
$p{\stackrel{a}{\to}}q$
 in $\hat M$, as we have constructed earlier, is the result of  two edges in $G_M$:
$p{\stackrel{w(a)}{\to}}\langle p, a, q\rangle {\stackrel{0}{\to} q  }$. Notice that
the algorithm has to assign probability 1 to the edge $\langle p, a, q\rangle {\stackrel{0}{\to} q  }$ since the outdegree of the node $\langle p, a, q\rangle$ is 1. Therefore, a random walk in $\hat M$ uniquely corresponds to a random walk
in $G_M$, while the only difference is the length is shortened by half. Therefore,
the random walk defined in $\hat M$ does achieve the rate $2\ln
\lambda_\diamondsuit$.  This  is also the fact even when the walk in $\hat M$ starts from the initial state and ends with an accepting state (as we did in the probabilistic program version of the $\hat M$), because of our earlier result in (\ref{earlier}).
Hence, once we prove the claim in (\ref{good}), the random words generated from
the probabilistic program defined by $\hat M$ do achieve the maximal free energy
in the Variational Principle.

 We now prove the claim in (\ref{good}).
Let $m>-\infty$ be the minimal weight of all transitions in $M$.
Consider
$$ \limsup \frac{1}{n} \ln{ \sum_{|\alpha|=n, \alpha \in L}  e^{w(\alpha)- mn} }=\lambda - m,  $$
where $\lambda=\limsup_{n \to \infty} \frac{1}{n} \ln{ \sum_{|\alpha|=n, \alpha \in L}  e^{w(\alpha)} } $ is defined in (\ref{deflambda}).
Note that the $\lambda\ge -\infty$.
(When $L$ is a finite language, the $\lambda=-\infty$. Otherwise,
the $\lambda$ is finite and $\lambda\ge m$.)
Fix a small $\epsilon>0.$ Then, by definition, there is an
$N_{\epsilon}>0$ such that
$\forall n >N_{\epsilon},$
$$E^{w(L_n) -nm}:= \sum_{|\alpha|=n, \alpha \in L}  e^{w(\alpha) -nm} \le e^{n(\lambda -m+\epsilon)},$$
where $L_n$ is the set of words in $L$ with length $n$. Notice that the
term $E^{w(L_n) -nm}$ is defined as 0 when $L_n=\emptyset$.
We now consider $M_{\diamondsuit}$ with the negative number
$w(\diamondsuit)$ satisfying that
${\hat \epsilon}=e^{w(\diamondsuit)}$ will make the following two items true:
\begin{itemize}
\item For each $l \le N_{\epsilon}$,
\begin{equation}\label{c1}
E^{w(L_{l})} {\hat \epsilon} \le \epsilon\epsilon^{l+1},
\end{equation}
where
$E^{w(L_{l})}:=\sum_{|\alpha|=n, \alpha \in L}  e^{w(\alpha)}$
 and therefore,
\begin{equation}\label{c2}
 E^{w(L_{l})- ml} {\hat \epsilon} \le \epsilon
\end{equation}
is also true.
\item ${\hat \epsilon} \le \epsilon$
\end{itemize}
Clearly,
  $M_{\diamondsuit}$  accepts a new language
$$ L^{\diamondsuit} := (L \diamondsuit)^{*} L $$
whose energy rate is, by definition in (\ref{defdimond}),
$$ \lambda_{\diamondsuit} := \limsup \frac{1}{n} \ln{\sum_{|\alpha|=n, \alpha \in L^{\diamondsuit}}} e^{w(\alpha)} $$
$$ = m + \limsup_{n \to \infty} \frac{1}{n} \ln{ \sum_{|\alpha|=n, \alpha \in L^{\diamondsuit}} e^{w(\alpha)-nm} }. $$
Notice that the $\lambda_\diamondsuit$ is a finite number since $L^\diamondsuit$ is not
a finite language (as mentioned earlier, we do not consider the trivial case when $L=\emptyset$ and, furthermore, the null word is taken out of the $L$.).
Now we focus on estimating the term
$\sum_{|\alpha|=n, \alpha \in L^{\diamondsuit}} e^{w(\alpha)-nm}$
with $n>N_\epsilon$.
Notice that each $\alpha$ in $L^{\diamondsuit}$ with length $n$ takes
the following form:
$$ \underbrace{ \underline{\alpha_{e_1}} }_{l_1} \diamond  \underbrace{ \underline{\alpha_{e_2}} }_{l_2} \diamond  \underbrace{ \underline{\alpha_{e_3}} }_{l_3} \diamond \dots $$
where $\alpha$ contains, say, $k$ diamondsuits for some $0\le k\le n.$
Each non-diamondsuit block has its own length , say, $l_1, l_2, \dots, l_{k+1},$
with $l_1 + l_2 + l_3 + \dots + l_{k+1} = n-k.$
Notice that $l_i>0$ for each $i$, recalling that the null word is taken out of $L$.
In this case,
each block is either ``short"(i.e., the length $\le N_{\epsilon}$) or ``long" (i.e., the length
$> N_{\epsilon}$).

Suppose that, among the $k+1$ blocks of lengths
$l_1, l_2, \dots, l_{k+1}$ respectively,  there are
$r$ short ones and $R$ long ones with $r+R=k+1$.
Furthermore, the total length of the long ones is denoted by Longlength $\le n$.
Of course, when $L$ is a finite language, there are no long blocks (when $N_\epsilon$ is large enough).

It is left to the reader to verify
the following two
cases:

\noindent Case 1. $L$ is an infinite language and hence
$+\infty>\lambda -m \ge 0$.
In this case,  the term to be estimated
$$\sum_{|\alpha|=n, \alpha \in L^{\diamondsuit}} e^{ w(\alpha) - nm}$$
$$ =\sum_{0\le k\le n, l_1+\dots +l_{k+1} = n-k, l_i>0}  E^{w(L_{l_1}) -m l_1} \times {\hat \epsilon} \times E^{w(L_{l_2}-m l_2)} \times {\hat \epsilon} \times \dots
E^{w(L_{l_{k+1}})- m l_{k+1}}  $$
(Using (\ref{c2}))
$$ \le \sum_{0\le k\le n, l_1 + \dots + l_{k+1}=n-k, l_i>0}  \epsilon^{r} e^{{\rm LongLength} (\epsilon + \lambda-m)}  \epsilon^{R}$$
$$ \le \sum_{0\le k\le n, l_1+\dots+l_{k+1}=n-k, l_i>0} \epsilon^{k+1} e^{n (\epsilon + \lambda-m)} $$
$$ \le \sum_{0\le k\le n, l_1+\dots+l_{k+1}=n-k, l_i \ge 0} \epsilon^{k+1} e^{n (\epsilon + \lambda-m)}  $$
$$ =\sum_{k=0}^{n} {n \choose k} \epsilon^{k} \epsilon e^{n (\epsilon + \lambda -m)} $$
$$ =(1+\epsilon)^{n} \epsilon e^{n (\epsilon + \lambda -m)}. $$
Therefore,
$$\lambda_{\diamondsuit} = m + \limsup_{n \to \infty} \frac{1}{n} \ln{\sum_{|\alpha|=n, \alpha \in L^{\diamondsuit}} e^{w(\alpha) - mn} }  $$
$$ \le m + \limsup_{n \to \infty} \frac{1}{n} \ln{ (1+\epsilon)^n \epsilon e^{n (\epsilon+\lambda-m)} }$$
$$ = m + \ln(1+\epsilon)+\epsilon +\lambda - m $$
$$ = \ln(1+\epsilon)+\epsilon + \lambda.$$

Case 2. $L$ is a finite language and hence $\lambda=-\infty$. Noticing that
we do not have any long blocks when $N_\epsilon$ is big enough.
In this case,  we estimate the following term instead
$$\sum_{|\alpha|=n, \alpha \in L^{\diamondsuit}} e^{ w(\alpha)}$$
$$ =\sum_{0\le k\le n, l_1+\dots +l_{k+1} = n-k, l_i>0}  E^{w(L_{l_1})} \times {\hat \epsilon} \times E^{w(L_{l_2})} \times {\hat \epsilon} \times \dots
E^{w(L_{l_{k+1}})}  $$
(because all blocks are short, we have the following, using (\ref{c1}))
$$ \le \sum_{0\le k\le n, l_1+\dots+l_{k+1}=n-k, l_i>0} \epsilon^n\epsilon^{k+1}$$
$$ \le \sum_{0\le k\le n, l_1+\dots+l_{k+1}=n-k, l_i \ge 0} \epsilon^n\epsilon^{k+1} $$
$$ =\sum_{k=0}^{n} {n \choose k} \epsilon^{k} \epsilon^n $$
$$ =(\epsilon(1+\epsilon))^{n}. $$
In this case,
therefore,
$$\lambda_{\diamondsuit} = \limsup_{n \to \infty} \frac{1}{n} \ln{\sum_{|\alpha|=n, \alpha \in L^{\diamondsuit}} e^{w(\alpha)} }  $$
$$ \le \limsup_{n \to \infty} \frac{1}{n} \ln{ (\epsilon(1+\epsilon))^n }$$
$$ =  \ln  \epsilon(1+\epsilon).$$

In Case 1,  $\lambda_\diamondsuit\to\lambda$ as $\epsilon\to 0$ (and hence $w(\diamondsuit)\to-\infty)$.
In Case 2, $\lambda_\diamondsuit\to\lambda=-\infty$ as $\epsilon\to 0$ (and, also, hence $w(\diamondsuit)\to-\infty)$.
That is, the claim in (\ref{good}) is valid since
$\lambda_{\diamondsuit}$ is monotonic in $w(\diamondsuit)$ according to the definition
in (\ref{defdimond}).

The Gurevich's
 Algorithm (\ref{Gurevich-algorithm}) is known efficient, numerically.
However, the translating from a regular language to a DFA is not when we use
textbook algorithms \cite{HopcroftU79}.  In fact, many online tools are
available for the translation and in many practical cases, it does not seem
terribly inefficient. Of course, it is meaningful future research to, efficiently,
 construct
 a random
walk program directly from an NFA accepting the $L$ while achieving
a reasonably good energy rate that is close to the $\lambda$.

\section{AEP for  a (Weighted) Regular Language}
Asymptotic Equipartition Property (AEP) is defined in Shannon's information theory to say that  sequences generated from certain random sources, with
probability approaching to 1, share roughly the same information rate (that is the rate of the source).
In literature, those sequences whose rate is close to the rate of the source
 are called typical sequences.
In this section,
we generalize the concept to a weighted regular language while the source
is the Markov chain $\hat M$ constructed from the given regular language $L$
in the previous section. Because the $\hat M$ defines the unique
$\mu^*$ in the Variational Principle, we can then define typical words in a weighted regular language.
Surprisingly, among all sequences, only a small number of sequences are typical even though they will show up with probability close to 1. This also says that, if we understand that a finite automaton is a probabilistic device to generate
words in a regular language by assigning transition probabilities  on the state
transitions and that the probability assignments actually make the device achieve Gurevich free energy,  then, with probability asymptotically 1, the device
will only generate a ``small" number of words, and each such word is typical.
If we understand that the regular language $L$ is used to specify a person's
normal behavior, then typical behavior
only takes a small portion of the $L$ --- most behavior  in $L$ is not typical.
Identifying such typical words has a lot of immediate applications in computer science, such as the ones listed below.
\begin{itemize}
\item In software testing, we may use the above idea to build two kinds of test suites: typical test suite and non-typical test suite.
Typical test suite is used to verify the programs implement basic and commonly used functionalities correctly.
Non-typical test suite creates challenges testing cases to check whether programs work correctly in the extreme cases.
\item In computer security, the typical words concept are similar to the Pareto principle (a.k.a. 80-20 rule).
Most behaviors are normal behaviors and the types of normal behaviors only take a small portion of all behaviors types.
While the abnormal behaviors occur rarely, the number of types regarding abnormal behaviors is much larger than that of normal ones.
Thus, we may use typical words to improve existing intrusion detection systems.
\end{itemize}

We start with claiming AEP on
aperiodic (i.e., there are two (nested) loops at the same node who lengths are co-prime)
and strongly connected
graphs.
Then, we apply the claim on weighted regular languages.

\subsection{AEP on a weighted, aperiodic and
 strongly connected graph}\label{aep}
Let $G$ be a weighted graph. We assume that $G$ is aperiodic
and is an SCC.
Let Gurevich Markov chain $\mu^*$
obtained from running
Algorithm \ref{Gurevich-algorithm} on the   $G$ be $p(\cdot)$, where
for each edge $t$ in $G$, $p(t)$ is the transition probability of $t$, and
$w(t)$ is the given
weight associated with $t$. Since the obtained $p(\cdot)$ is ergodic,
we
assume that $\pi(\cdot)$ is the unique
stationary distribution where $\pi(v)$ is the stationary
probability on node $v$ in $G$, for each node $v$.
We shall use $\pi(t)$ to denote $\pi(v)$ where $v$
is the starting node of the edge $t$.
A walk $\alpha$ of $G$ is a sequence of edges
\begin{equation}\label{walk}
\alpha=t_1\cdots t_n
\end{equation}
 for some $n$ such that
the ending node of $t_i$ equals the starting node of $t_{i+1}$ for each $i$.
For the walk $\alpha$,
we use $P(\alpha)$
to denote the probability that $\alpha$ is actually walked; i.e.,
$P(\alpha)=\pi(t_1) p(t_1) \cdots p(t_n)$.

Let $\epsilon>0$.
 We say that the walk $\alpha$ in (\ref{walk})
 is $\epsilon$-typical (of $G$), if
$$|\lambda_{\alpha} - \lambda_{G}| \le \epsilon,$$
where $\lambda_{G}$ (recalling that $\lambda_{G}$
 is the $\ln\hat\lambda$ where
$\hat\lambda$ is the Perron number computed in Algorithm \ref{Gurevich-algorithm})
 is the free energy rate of $G$,
and $\lambda_{\alpha}$ is the free energy rate of the walk $\alpha$; i.e.,
$$ \lambda_{\alpha} = \frac{\sum_{i=1}^{n} w(t_i) - \ln p(t_i) }{|\alpha|}. $$
We define $T_{\epsilon, G}^{n}$ to be set of $\epsilon$-typical walks
in $G$ with length $n$. Then, we claim that
$$\lim_{n \to \infty} \sum_{\alpha \in T_{\epsilon, G}^{n}} P(\alpha)  =1,$$
where $p$ is  defined above from the Gurevich Markov chain on $G$. Its proof is as follows.
\begin{proof}
Notice that the Gurevich Markov chain $p(t)$ where $t$ is an edge in $G$,
is ergodic, and hence, for any fixed $n$, we use $X_n$
to denote a random variable over all walks $t_1 t_2 \dots t_n$ on $G$ with
length $n$.
Then,
$$ {\bf E}\{ \lambda_{X_n} \}  = \sum_{t_1 \dots t_n \text{ is a walk}} \frac{\sum_{i=1}^{n} w(t_i) - \ln p(t_i)  }{n}  P(t_1 \dots t_n)$$
$$ = \frac{1}{n} \sum_{t_1 \dots t_n \text{ is a walk}} \sum_{i=1}^n [w(t_i) - \ln p(t_i)]\pi(t_1) p(t_1) \dots p(t_n) $$
$$=\frac{1}{n} \sum_{t_1 \dots t_n \text{ is a walk}} [w(t_1) -\ln p(t_1)
]\pi(t_1) p(t_1)
\dots p(t_n) +\cdots$$
$$+ \frac{1}{n} \sum_{t_1 \dots t_n  \text{ is a walk}  }  [w(t_n) -\ln{p(t_n})] \pi(t_1) p(t_1) \dots p(t_n) $$
$$ = \frac{1}{n} \sum_{t\text{ is an edge}}
  \sum_{\substack{  t_1=t\\ t_1 \dots t_n  \text{ is a walk}} }
[w(t_1) -\ln p(t_1)
]\pi(t_1) p(t_1)
\dots p(t_n) +\cdots$$
$$ + \frac{1}{n} \sum_{t\text{ is an edge}}
  \sum_{\substack{  t_n=t\\ t_1 \dots t_n  \text{ is a walk}} }
[w(t_n) -\ln p(t_n)
]\pi(t_1) p(t_1)
\dots p(t_n)$$
(Using the fact that $\pi(\cdot)$ is the stationary distribution)
$$ = \frac{1}{n} \sum_{t \text{ is an edge}} [w(t) - \ln {p(t}] \pi(t) p(t) +\cdots$$
$$ + \frac{1}{n} \sum_{t \text{ is an edge}}  [w(t)-\ln {p(t})] \pi(t) p(t)$$
$$ = \sum_{t \text{ is an edge}} w(t) \pi(t) p(t) - \pi(t) p(t) \ln{p(t)}  $$
(Using the RHS of the Variational Principle)
$$ = \lambda_G $$

\noindent Hence, from the  law of large numbers for ergodic Markov chain,
we have
$$ \lim_{n \to \infty} \text{Prob} \{  |\lambda_{X_n} - \lambda_{G}| \le \epsilon \} =1.$$
The claim follows.
\end{proof}

\subsection{AEP for a weighted regular language}\label{cluster}
Let $L$ be a weighted regular language. In the construction presented
 in Section
\ref{section2}, the strongly connected
graph $G_M$ is run on
Gurevich's  Algorithm \ref{Gurevich-algorithm}.
However, in order to use the AEP results in Section \ref{aep},
the graph must be aperiodic.
This can be easily resolved as follows.

\begin{figure}
\begin{tikzpicture}[->,>=stealth',shorten >=1pt,auto,node distance=2.1cm,
        scale = 1,transform shape]

  \node[state] (q0) {$q_0$};
 \node[state] (q01) [right of=q0] { $q_{0a1}$};
  \node[state] (q1) [right of=q01] {$q_1$};
\node[state] (q10) [above of=q01] {$q_{1a0}$};
\node[state] (q11) [above of=q1] {$q_{1b1}$};

 \node[state](F)[below of =q1]{$F$};
\node[state](clubsuitq1)[right of =q1]{$\clubsuit_{q_1}$};

  \path
(clubsuitq1)edge         node {$w(\spadesuit)=0$} (F)
(q1)edge         node {$w(\clubsuit)=0$} (clubsuitq1)
(q1)edge         node {$w(\heartsuit)=0$} (F)
(F)edge         node {$w(\diamondsuit)=-1000$} (q0)
(q0) edge              node {$w(a)$} (q01)
         (q01) edge              node {$0$} (q1)
        (q1) edge           node {$w(a)$} (q10)
(q10) edge              node {0} (q0)
        (q1) edge    [bend left]          node {$w(b)$} (q11)
  (q11) edge    [bend left]          node {$0$} (q1);

\end{tikzpicture}
\caption{The weighted graph $G_{\tilde M}$ converted from $\tilde M$.}\label{exdfa9}
\end{figure}
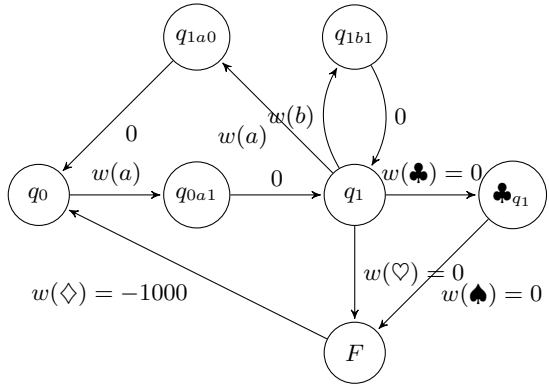

We now consider a new language $\tilde L= L(\clubsuit\spadesuit+\heartsuit)$
where $\clubsuit,\spadesuit,\heartsuit$ are new symbols with weight 0.
Suppose that $M$ is the minimal DFA  accepting $L$. We now
construct $\tilde M$ by modifying the $M$ as follows
\begin{itemize}
\item Add a new state named $F$;
\item For each accepting state $q$ of $M$, add a state $\clubsuit_q$ as well
as three transitions
$q\xrightarrow{\clubsuit}\clubsuit_q$, $\clubsuit_q\xrightarrow{\spadesuit}F$,
$q\xrightarrow{\heartsuit}F$;
\item Make $F$ be the only accepting state of $\tilde M$.
\end{itemize}
In Section
\ref{section2},  the strongly connected graph $G_M$ is constructed by
splitting every transition in $M$ into two transitions (see Figure \ref{exdfa3}).
We can construct $G_{\tilde M}$ similarly but without splitting all the
newly added edges in $\hat M$. For instance, for the example DFA $M$
in Figure \ref{exdfa}, the $G_{\tilde M}$ is shown in
Figure \ref{exdfa9}, which can be compared with Figure \ref{exdfa3} where
 the  difference can be seen.

Clearly, $G_{\tilde M}$ is aperiodic.


From $\tilde L$, the
graph $G_{\tilde M}$ is constructed accordingly. Consequently, as described in Section \ref{section2},
a Markov chain $\mu^*$ that assigns transition probability $p(t)$ to an
edge $t$ in the graph $G_{\tilde M}$ is
computed using Algorithm \ref{Gurevich-algorithm}.
We now first recall the following the definition in Section \ref{aep}.
Let $K$ be a (large) number and $\epsilon>0$ be a small number.
Notice that, herein,  the (negative) weight $w(\diamond)$ in $G_{\tilde M}$ gives $\hat\epsilon=e^{w(\diamond)}$.
A walk $T$ in $G_{\tilde M}$
with length $K$ is {\em $(\epsilon,\hat\epsilon)$-typical} if
\begin{equation}\label{taep}
|\lambda_{G_{\tilde M}}-{1\over K  }\sum_{1\le i\le k} w(t_i)-\ln p(t_i) |<\epsilon,
\end{equation}
where $\lambda_{G_{\tilde M}}$ the free energy rate of the graph $G_{\tilde M}$.
Though $G_{\tilde M}$ is a graph (without ``initial" and ``accepting" nodes), the initial/accepting state in $M$
can be uniquely located in $G_{\tilde M}$ through the construction of $G_{\tilde M}$. Hence, we can still
say that a walk $T$ is initial when it starts with the initial node that corresponds to the initial state of $M$.
Since the Markov chain $\mu^*$ computed on the graph $G_{\tilde M}$ is ergordic, and hence,
\begin{equation}
\lim_{K\to\infty} {\rm Prob~} \{ T: T \text{ is an initial walk with length } K\}= 1.
\end{equation}
Using the AEP result in Section \ref{aep}, we then have
\begin{equation}\label{lim1}
\lim_{K\to\infty} {\rm Prob~} \{ T: T \text{ is an initial  and }(\epsilon,\hat\epsilon)\text{-typical walk with length } K\}= 1.
\end{equation}

For each initial and $(\epsilon,\hat\epsilon)$-typical walk $T$, one can uniquely recover, by reversing the construction
from $M$ all the way to the graph $G_{\tilde M}$,
a word $\sigma(T)$ that is a prefix of a word in
 $(\tilde L\diamondsuit)^*\tilde L$. Noticing that the $\alpha$ is in the form of
$$\alpha_1\#\cdots \#\alpha_k$$
for some $k$, where each $\alpha_i\in L$ (except for the last $\alpha_k$, see below)
 and the
$\#$ is either $\clubsuit\spadesuit\diamondsuit$ or
$\heartsuit\diamondsuit$.  The last $\alpha_k$, when  ended with an
auxiliary  symbol (like $\clubsuit, \spadesuit, \diamondsuit$), we can simply remove those
symbols and obtain a word (also denoted as $\alpha_k$) in $L$.
 When, however, $\alpha_k$ is not ended with an auxiliary  symbol, we may append the $\alpha_k$ with
a bounded length word so that the resulting word (still denoted as $\alpha_k$) will be in $L$.
In this latter case, one can think the $T$ is appended with a few more edges and the resulting $T$'s typicalness will not be affected if the length $K$  of $T$ is taken large enough.

We shall now say that the {\em cluster} (a multiset)  of words in $L$
$$\{\alpha_1,\cdots,\alpha_k\}$$
obtained from the $T$ is $(\epsilon,\hat\epsilon, K)$-typical in $L$ when the $T$  of length $K$
is $(\epsilon,\hat\epsilon)$-typical.
We use the cluster instead of the sequence $\alpha_1,\cdots,\alpha_k$ here since
the ordering of the appearance of the $\alpha_i$'s in $\sigma(T)$ has no effect for the typicalness
(it is left to the reader to verify).

Surprisingly, computing an $(\epsilon,\hat\epsilon, K)$-typical cluster  in $L$ is not as obvious as it looks.
A direct implementation would include the following steps:
\begin{enumerate}
\item[1] Construct the $M$ (minimal DFA for $L$);
\item[2] Exercise all the steps in Section \ref{section2} to obtain $G_{\tilde M}$;
\item[3] Run Algorithm \ref{Gurevich-algorithm} on $G_{\tilde M}$ so that transition probabilities can be assigned to the edges in  $G_{\tilde M}$ and, at the same time, obtain the free energy rate $\lambda_{G_{\tilde M}}$;
\item[4]  Find an initial walk $T$ (which may contain loops and nested loops) on the graph $G_{\tilde M}$
so that the constraint in (\ref{taep}) is satisfied;
\item[5] We recover and return a cluster constructed from the $T$ as described in above.
\end{enumerate}

\noindent Though Step 1 is, in theory, not efficient. However, for practical purpose, as we have
mentioned at the end of
Section \ref{section2}, it may not be bad at all using some well established online tools.
The main difficulty is, however, Step 4. This is because all the initial walks $T$
that we are looking for  form a semilinear language
whose Parikh map is not so easy to compute. One possible way is sketched as follows.
One can construct a system of integer linear/modular constraints, through loop analysis,
 to characterize the semilinear set (on edge counts) of all the initial
walks on the graph (see, e.g., the paper \cite{XieLD04}, and then,
translate the constraint in (\ref{taep}) into linear constraints on the counts of individual edge counts, and finally,
an integer linear constrain solver may be used to obtain a solution (as a vector of
 counts of edges). One final step would be
to recover the vector to a walk $T$. This whole process is terribly inefficient and not practical at all.
Unfortunately, this is the best algorithm we can come up with, for now, since the problem in Step 4
is essentially asking an algorithm to decide whether there is a walk on a graph whose vector of edge counts
satisfies a given linear constraint system.  In \cite{CuiDF11}, we investigated ``typicalness"
in a setting where we used branching factor of a graph to approximate information rate (instead of
energy rate)  of the graph, where a model-checker SPIN \cite{holzmann1997model} was used to identify a typical path
of a graph. This model-checking approach might serve an alternative way to resolve Step 4, practically.

Luckily, we shall notice that, as shown in (\ref{lim1}), for a large $K$, the initial and typical walks take probability asymptotically 1. That is, if we just randomly walk on the Markov chain $\mu^*$ assigned on the graph
$G_{\tilde M}$, we will have probability close to 1 to obtain a desired $T$. Hence, the following random algorithm
is a far more practical and efficient solution to Step 4:
\begin{enumerate}
\item[4.1] Assign transition probabilities to edges in $G_{\tilde M}$ according to the computed Markov chain $\mu^*$ in Step 3;
\item[4.2] Treat the graph now as a probabilistic program and by running the program from the initial node, we create a random walk of length $T$;
\item[4.3] If $T$ satisfies (\ref{taep}), return, else goto Step 4.2;
\end{enumerate}
One should be able to obtain the expected number of rounds on the loop between Step 4.2 and Step 4.3, though analyzing the mixing time of the ergordic Markov chain $\mu^*$, but we leave this for future work.

\subsection{Special cases when a typical cluster is a singleton}\label{singleton}
One shall notice that, through out the paper, if
the  minimal DFA $M$ of  the  regular language $L$ is already strongly connected
and aperiodic, we do not need any $\diamondsuit/\heartsuit/\clubsuit$-edges in the
$G_{\tilde M}$ and hence, the $\hat\epsilon$-edges are not needed and
consequently, an obtained typical cluster will always contain one word (i.e., the cluster is a singleton);
in this special case, we can
say that the word is typical in $L$. Such a regular language would inherently a ``looped" language
such as $(ab+c)^*$.

There is another special case
when $L$ is prefix-closed that is very  useful in practice. In this case, every state in
the minimal DFA $M$ accepting $L$ is an accepting state.
Such a prefix-closed $L$ can be used to specify applications like network traffic (a prefix of the traffic stream is also
a traffic stream).

Recall that, in Section \ref{aep},
the AEP is established on the Markov chain
$G_{\tilde M}$ (with transition probabilities in $\mu^*$)
constructed from the regular language $L$. We assume that $L$ is an infinite language
(and therefore its energy rate $\lambda$ is finite)
and hence the energy rate
$\lambda_\diamondsuit$ of $G_{\tilde M}$ approaches the finite number
$\lambda$ as $w(\diamondsuit)\to-\infty$. In Section \ref{cluster}, we use the
notation $\sigma(T)$ to denote the unique word ``recovered" from a walk
on the graph $G_{\tilde M}$.  The walk $T$ is called {\em accepting} if $T$ starts with $q_0$ and
ends with the node $F$, where in between,  it never passes the node $F$.  Such an accepting  walk $T$
 can also be understood as the following: the Markov chain
starts with $q_0$ and walks along the $T$. When it ends at the  node $F$, it is also the {\em first hitting time}
(the length of $T$) for the node $F$. We use $\tau$ to denote the random variable where
${\rm Prob} \{\tau=n\}$ is the probability that there is an accepting walk $T$ with length $n$.
There are some existing results on the distribution of $\tau$.
For instance, from Maciuca and Zhu \cite{maciuca2006first},
explicit formulas are given (see formulas (2.2) in the paper)
on ${\bf E} (\tau)$ through the fundamental matrix of the
Markov chain. Notice that, herein, the $\tau$ is for the Markov chain starting from
the node $q_0$ instead of starting from the unique stationary distribution
$\pi(\cdot)$.  The $\tau$ can be approximated well with an exponential distribution
(Prop 3.7 in \cite{maciuca2006first}). Hence, with high probability,
the $\tau$ can not be too bigger than the mean ${\bf E} (\tau)$.
Even though it is still difficult to estimate the value of ${\bf E} (\tau)$,
we can have a clue of it from the well known result:
when the Markov chain starts from the unique stationary distribution $\pi(\cdot)$,
${\bf E} (\tau)={1\over \pi(F)}$. Furthermore,
$|\pi(F)\cdot w(\diamondsuit)|<c$ for some constant $c$ when $w(\diamondsuit)\to -\infty$.
(This can be shown by the fact $\lambda_\diamond\to\lambda>-\infty$ as $w(\diamondsuit)\to -\infty$
and by the ergordicity theorem of the Markov chain; we omit the details.)
Therefore, ${\bf E} (\tau)$ should be at the same magnitude of $-w(\diamondsuit)$
(which is big according to our choice of $w(\diamondsuit)$).
Therefore, for a large $N$ (but less than $O(-w(\diamondsuit))$), the set $S_N$
of all initial (i.e., starting from $q_0$)
walks $T$ (with length $N$)
 that is a prefix of an accepting walk should take a high probability (among all initial walks with length $N$)
since, as we have said, the probability that $N<\tau$ is high. Notice that all initial walks with length $N$
are with probability close to 1 (when the $N$ is large), and also from the AEP in Section \ref{aep},
we shall expect the following:
all $(\epsilon,\hat\epsilon,N)$-typical walks $T$ in the $S_N$ take a high probability.
Notice that since $L$ is prefix closed, each such typical $T$ uniquely corresponds to a word
in $L$ and hence each such word can be called typical.
Of course, the above analysis is very informal   since it is very difficult to
make a precise estimate on the hitting times. But, we think this shall good enough for a practitioner to
successively (with high probability  --- that is $1-\delta$ where the small
$\delta$ depends on $\epsilon, w(\diamondsuit)$ but {\em not} on $N$, and hence it is merely
an approximation of the
AEP)
generate a long typical word from the prefix-closed $L$ as follows:
(on given $\epsilon, \hat\epsilon=e^{w(\diamondsuit)}, N$)

Step 1. Construct the Markov chain $G_{\tilde M}$ (with transition probabilities in $\mu^*$), from the minimal automaton $M$ for the $L$;

Step 2. Treat $G_{\tilde M}$ as a probabilistic program and starts to walk from the  node $q_0$;

Step 3. When the program hits state $F$, goto Step 2.  When the program hits the length $N$,
if the walk so far  is indeed $(\epsilon,\hat\epsilon, N)$-typical
(using its definition in Section \ref{aep}), then return the word (in $L$) recovered from the walk
as a typical word, else goto Step 2.

\subsection{Discussions on a typical cluster in a weighted regular language}\label{nomodel}
What are the practical implications for a typical cluster? One can interpret it in the following way.
Suppose that we are allowed to pick words from a bag $L$ of words. This bag can be finite or infinite.
It is required to pick words whose total length is (roughly) $K$ that is a large given number. What would be
the most typical way to pick? A typical cluster obtained in Section \ref{cluster} fits this purpose.

One can also re-interpret the cluster under various context. For instance,
given a software under test, one may generate a test suite (a set of test cases, where each test case is simply
an input sequence) from a given requirements specification which can be a regular language (on input/output
events) or a finite automaton to accept the language. Then, what would be a typical test suite? (and according,
what would be a non-typical test suite?) Notice that weights can be naturally assigned with each event to indicate
for instance the cost of running a test on the event.
 A typical cluster obtained in
Section \ref{cluster} may also serve this purpose.
One other example is abnormal detection in computer security or in social security.
Suppose that $A$ is a device (such as a network, a medical monitor, etc.,)  or
a malicious person that we already have a specification of $A$'s ``normal" behavior.
Such a specification can be simply a finite automaton or a regular language, to define normal or
 legal sequences
of activities. Each activity can itself associated with some physical quantity such as time, money spent,
or even risk level.
However, a difficulty in computer security is to identify abnormal in normal. This is because an abnormal
behavior can be completely legal but sometimes, it is a prelude to a terrible attack.
(For instance, a person circulating a police office three times a day is legal but abnormal.)
It is often the case that a set of behaviors (i.e., sequences of activities) are obtained through, for instance,
probing a network traffic, or targeted surveillance. Consequently, we may apply the algorithms in
Section \ref{cluster} to decide whether such a set is typical or non-typical, which has great importance in
practice.

Finally, we would like to add a few words on the  abnormal detection in normal mentioned earlier.
In reality, it is very difficult to give a specification for normal; in many cases the specification in partial,
error-prone, or
even missing.  In the case that without a specification, can we still detect normal/abnormal?  In other words,
can we solve the following problem:
\begin{enumerate}
\item[]  Given:  a number of  large
 finite sets $S_1,\cdots, S_k$ (some or all of the sets can be singletons, see Section \ref{singleton})
 of weighted activity sequences (on a known weighted activity
alphabet $\Sigma$);
\item[]  Goal:  identify those sets that are typical/non-typical (i.e., normal/abnormal).
\end{enumerate}
Notice that, herein, there is no automaton or regular language given. Our solutions are as follows.

Step a. For each set $S_i$, every sequence
 in $S_i$ obtain a random but unique id $\eta_{ij}$ in $[1, |S_i|]$.
Now, we use the $\beta_{ij}$ to represent the sequence whose id is
$\eta_{ij}$. Notice that each activity symbol
in $\beta_{ij}$ is weighted.
Then, we concatenate all the sequences
 $\beta_{ij}$  to a new long sequence $\beta_i$. (For example,
suppose that the set $S_1$ has three sequence $abc$, $def$, and $hij$.
Let $\beta_{11}=def, \beta_{12}=hij$ and $\beta_{13}=abc$. Then, $\beta_1=defhijabc$)

Step b. For each $\beta_i$, for every symbol $a$, we count its occurrences in $\beta_i$, $\#_{\beta_i}(a)$.
Then, estimate the probability of symbol $a$, i.e., $Pr(a)= \frac{\#_{\beta_i}(a)}{|\beta_i|} $ .
So, we have $\int w d\mu \approx \sum_{ a} \frac{\#_{\beta_i}(a)}{|\beta_i|}$.

Step c.  Run Lempel-Ziv compression algorithm (such as the one implemented in {\tt 7z}
\cite{7zwebsite}) on each $\beta_i$ so that we obtain
the reciprocal of the compression ratio which  is an estimation
of the entropy $H_{\beta_i}$ (converted in natural logorithm).
Then, we have $\lambda_i = \sum_{ a \in \Sigma_{\beta_i}} \frac{\#_{\beta_i}(a)}{|\beta_i|} + H_{\beta_i}$;

Step c. Compute $\lambda={1\over k}\sum\lambda_i$;

Step d. For the given $\epsilon$ and for each $i$,
if $|\lambda-\lambda_i|<\epsilon$, report that $S_i$ is typical/normal, else
report  that $S_i$ is non-typical/abnormal.

\noindent Notice that the reason why this approach works is due to the fact that Lempel-Ziv can be used to
approach the free energy rate after the transcoding in Step a is applied. However, it does need a side condition
(for Lempel-Ziv algorithm):
the source that generate the sequences must be stationary and ergodic, which is the same as the constraints
that our graph $G_{\tilde M}$ satisfies. We leave this in the journal version of the paper for a  more detailed presentation.

Of course, the nontypical/abnormal detection algorithms we presented so far on the case that the language or the automaton is presented and on the case that the language or the automaton is missing also give a way to
clustering weighted stream data with a model or without a model (unsupervised). Currently, our PhD student
William Hutton is conducting experiments on using the algorithms to detect
abnormal
TCP/IP handshaking traffic for the case when a model is present and the case when the model  is not available.
In the future, we would like to generalize the experiments to a broader setting so that more practical effectiveness
results can be obtained. For the completeness of this paper, we present a small but real-word example to see
how one would use the algorithms in software testing.

\section{Applications: typical and non-typical test cases}
Model-based software testing generates test codes from a labeled
transition system, served as the model of the system under test.
ModelJunit \cite{modeljunit} is one such testing tool in which one can specify
 a labeled transition system and the tool
may also produce test cases automatically.
We now look at an example, which is taken from an online  presentation \cite{example}.

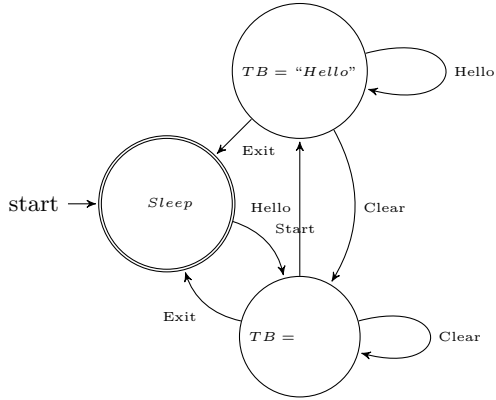
\begin{figure}[h]
\begin{tikzpicture}[->,>=stealth',shorten >=1pt,auto,node distance=2.5cm,
        scale = 1,transform shape]

  \node[state,initial, accepting] (sleep) {\tiny$ {~~~~~~Sleep~~~~~}$};
  \node[state] (hello) [above right of=sleep] {\tiny$TB=``Hello"$};
  \node[state] (space) [below right of=sleep] {\tiny$TB =~~~~~~~~$};
  \path (sleep) edge   [bend left]           node {\tiny Start} (space)
        (hello) edge       [loop right]        node {\tiny Hello} (hello)
(space) edge       [loop right]        node {\tiny Clear} (space)
      (hello) edge              node {\tiny Exit} (sleep)
        (space) edge     [bend left]          node {\tiny Exit} (sleep)
        (hello) edge          [bend left]       node {\tiny Clear} (space)
        (space) edge              node {\tiny Hello} (hello);

\end{tikzpicture}
\end{figure}

This transition system is intended to model an implementation of a simple
text box where a user can fill in ``Hello" or clear the text in
the box. A test sequence is a sequence of events on the user inputs, which
is the sequence of labels on a walk on the graph shown above.
To ease our presentation, we take the state ``sleep" as the accepting
state. (So, we consider a complete cycle of the implementation.)
We call the transition system as the DFA $M$.
  Notice that the DFA is already
minimal, strongly connected and aperiodic. Hence, following the discussions at the beginning of Section \ref{singleton}, we can directly compute the transition probabilities, using the Markov chain
that achieves the free energy rate of $L(M)$,
 on transitions in $M$ without
introducing any $\hat\epsilon$-edges, as shown in the following figure, which is a probabilistic program
denoted as $\hat M$ (
Here, we implicitly assume that the input labels share the same weight (i.e., = 1).
):

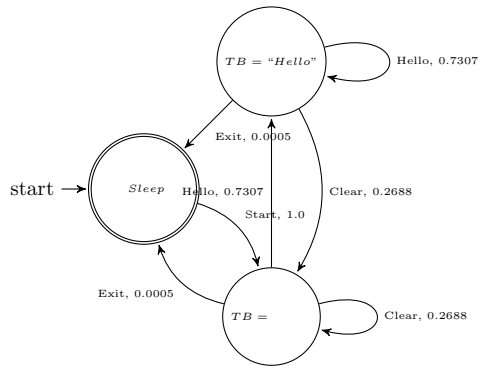
\begin{tikzpicture}[->,>=stealth',shorten >=1pt,auto,node distance=3.0cm,
        scale = 0.8,transform shape]

  \node[state,initial, accepting] (sleep) {\tiny$ {~~~~~~Sleep~~~~~}$};
  \node[state] (hello) [above right of=sleep] {\tiny$TB=``Hello"$};
  \node[state] (space) [below right of=sleep] {\tiny$TB =~~~~~~~~$};
  \path (sleep) edge   [bend left]              node {\tiny Start, 1.0} (space)
        (hello) edge       [loop right]         node {\tiny Hello, 0.7307} (hello)
        (space) edge       [loop right]         node {\tiny Clear, 0.2688} (space)
        (hello) edge                            node {\tiny Exit, 0.0005} (sleep)
        (space) edge     [bend left]            node {\tiny Exit, 0.0005} (sleep)
        (hello) edge          [bend left]       node {\tiny Clear, 0.2688} (space)
        (space) edge              node {\tiny Hello, 0.7307} (hello);

\end{tikzpicture}


From this, we can see the test cases can be generated automatically when
the $\hat M$ is run as a probabilistic program.

Notice that, not all test cases are born equal. From the AEP theorem, the program,
asymptotically with probability 1, will generate a test case that is always
typical. However, there are indeed nontypical test cases, which are also valuable
(i.e., those test cases may reveal a fault that is not possible to detect when the
system under test is exercised ``normally"). Therefore, a ``good" test suite
shall not only include typical test cases, but also include nontypical test cases.
We now look at the following example test suite.
$T_1$ is the set of the following four test cases.

$Test_1 = Start Helllo Exit$,

$Test_2 = Start Hello Clear Exit $,

$Test_3 = Start Clear Exit $,

$Test_4 = Start Hello Hello Exit $.

\noindent Notice that, by walking each test case on the graph, every
branch(transition) is exercised at least one. Hence,
the suite $T_1$ achieves 100 \% branch-coverage.

We now take a small $\epsilon = 0.1$ and verify that
$Test_1, Test_2, Test_3, Test_4$ are all $\epsilon$-typical.
Next, we define the following suite $T_2$ of four test cases:

$ Test'_1 = Start Helllo Exit $,

$ Test'_2 = Start Exit Start Exit Start Exit $,

$ Test'_3 = Start Hello  HelloHelloHello Exit $,

$ Test'_4 = Start Hello  Clear Clear Exit$.

\noindent Again,
we can compute that
$Test'_1$, $Test'_3$ and $Test'_4$  are $\epsilon$-typical
but, $Test'_2$  is not $\epsilon$-typical.
Notice that $T_2$ also achieves 100\% branch coverage.
That is, $T_2$ may have a better chance to find
``corner faults".
Therefore, our approach can also be used to evaluate
an  existing test suite and see if it contains a reasonable portion
of typical test cases.  We shall also point out that the discussions made in Section \ref{nomodel}
can also be applied to identifying typical/non-typical test cases in a given test suite even when the model
is not given.

Finally, we shall point out that when the weights assigned to the input labels are changed, so are the test cases'
typicalness (since the free energy rate and the Markov chain are
accordingly changed). For instance, if we assign $w(Clear)=4, w(Hello)=5, w(Start)=1, w(Exit)=2$
(the weight of an input label
 can be used to measure, e.g., the cost associated with the run of the system under test when the input
label is executed.) and $\epsilon$ is the same as above, then we can verify that
the test case $Test'_3$ becomes typical.

\section{Conclusions}
We study random words in a weighted regular language  that achieve the maximal free energy using thermodynamics formalism. In particular, typical words in the language are algorithmically generated which have applications in
computer security (anomaly detection) and software enegineering (test case generation).
  In the future, we may continue the effort to apply our approaches to larger scale
real world
applications.

\bibliography{randomwords}{}

\begin{thebibliography}{10}

\bibitem{7zwebsite}
7zip.
\newblock \url{http://www.7-zip.org/}.

\bibitem{aldous2002reversible}
David Aldous and Jim Fill.
\newblock Reversible markov chains and random walks on graphs, 2002.

\bibitem{basset2015maximal}
Nicolas Basset.
\newblock A maximal entropy stochastic process for a timed automaton.
\newblock {\em Information and Computation}, 243:50--74, 2015.

\bibitem{CuiD16}
Cewei Cui and Zhe Dang.
\newblock A free energy foundation of semantic similarity in automata and
  languages.
\newblock In {\em Similarity Search and Applications - 9th International
  Conference, {SISAP} 2016, Tokyo, Japan, October 24-26, 2016. Proceedings},
  pages 34--47, 2016.

\bibitem{CuiDF11}
Cewei Cui, Zhe Dang, and Thomas~R. Fischer.
\newblock Typical paths of a graph.
\newblock {\em Fundam. Inform.}, 110(1-4):95--109, 2011.

\bibitem{CuiDFI16}
Cewei Cui, Zhe Dang, Thomas~R. Fischer, and Oscar~H. Ibarra.
\newblock Execution information rate for some classes of automata.
\newblock {\em Information and Computation}, 246:20--29, 2016.

\bibitem{CuiDFI17}
Cewei Cui, Zhe Dang, Thomas~R. Fischer, and Oscar~H. Ibarra.
\newblock Information rate of some classes of non-regular languages: An
  automata-theoretic approach.
\newblock {\em Information and Computation}, (to appear).

\bibitem{Gurevich1984}
B.~M. Gurevich.
\newblock A variational characterization of one-dimensional countable state
  gibbs random fields.
\newblock {\em Zeitschrift f{\"u}r Wahrscheinlichkeitstheorie und Verwandte
  Gebiete}, 68(2):205--242, 1984.

\bibitem{holzmann1997model}
Gerard~J. Holzmann.
\newblock The model checker spin.
\newblock {\em IEEE Transactions on software engineering}, 23(5):279--295,
  1997.

\bibitem{HopcroftU79}
John~E. Hopcroft and Jeffrey~D. Ullman.
\newblock {\em Introduction to Automata Theory, Languages and Computation}.
\newblock Addison-Wesley, 1979.

\bibitem{lovasz1993random}
Laszlo Lovasz.
\newblock Random walks on graphs: A survey.
\newblock {\em Combinatorics, Paul Erdos in Eighty}, 2, 1993.

\bibitem{maciuca2006first}
Romeo Maciuca and Song-Chun Zhu.
\newblock First hitting time analysis of the independence metropolis sampler.
\newblock {\em Journal of Theoretical Probability}, 19(1):235--261, 2006.

\bibitem{malkiel1999random}
Burton~Gordon Malkiel.
\newblock {\em A random walk down Wall Street: including a life-cycle guide to
  personal investing}.
\newblock WW Norton \& Company, 1999.

\bibitem{modeljunit}
ModelJunit.
\newblock \url{https://sourceforge.net/projects/modeljunit/}.

\bibitem{page1999pagerank}
Lawrence Page, Sergey Brin, Rajeev Motwani, and Terry Winograd.
\newblock The pagerank citation ranking: Bringing order to the web.
\newblock Technical report, Stanford InfoLab, 1999.

\bibitem{example}
Kyra~Hameleers Richid~Kherrazi and Adrian Iankov.
\newblock Hands-on experience model based testing with spec explorer.
\newblock
  \url{https://www.slideshare.net/Rachid99/handson-experience-model-based-testing-with-spec-explorer}.

\bibitem{ruelle2004}
D.~Ruelle.
\newblock {\em Thermodynamic Formalism: The Mathematical Structure of
  Equilibrium Statistical Mechanics}.
\newblock Cambridge Mathematical Library. Cambridge University Press, 2004.

\bibitem{sarig-notes}
Omri~M. Sarig.
\newblock Lecture notes on thermodynamic formalism for topological markov
  shifts, 2009.

\bibitem{shabook}
C.~E. Shannon and W.~Weaver.
\newblock {\em The Mathematical Theory of Communication}.
\newblock University of Illinois Press, 1949.

\bibitem{XieLD04}
Gaoyan Xie, Cheng Li, and Zhe Dang.
\newblock Linear reachability problems and minimal solutions to linear
  diophantine equation systems.
\newblock {\em Theor. Comput. Sci.}, 328(1-2):203--219, 2004.

\end{thebibliography}
\bibliographystyle{plain}

\end{document}